\documentclass[showpacs,preprintnumbers,prb]{revtex4}

\usepackage{graphicx}
\usepackage{amssymb}

\begin{document}

\title{Penetration of an external magnetic field into a multistrip superconductor/soft-magnet heterostructure}

\author{S.~V.~Yampolskii}
\email{yampolsk@tgm.tu-darmstadt.de}
\author{Yu.~A.~Genenko}
\author{H.~Rauh}
\affiliation{Institut f\"{u}r Materialwissenschaft, Technische
Universit\"{a}t Darmstadt, 64287 Darmstadt, Germany}
\date{28 June, 2006}

\begin{abstract}
The magnetization of a planar heterostructure of periodically alternating 
type-II superconductor and soft-magnet strips exposed to a transverse external magnetic field is studied. 
An integral equation governing the sheet current distribution in the Meissner state of 
the superconductor constituents is derived. The field of complete penetration of magnetic flux 
in the critical state of the superconductor constituents is calculated for different widths of the 
superconductor and the soft-magnet constituents and a range of values of the relative 
permeability of the soft-magnet constituents. 
\end{abstract}

\maketitle

Heterostructures made up of superconductor (SC) and soft-magnet (SM) constituents are 
being studied extensively because of their potential for improving the performance of SCs. 
The use of SMs for such structures offers the possibility to alter the pinning of magnetic 
vortices in the SC constituents through easy tuning of the intrinsic magnetic moment of 
the SM constituents~\cite{SMdot1,SMdot2}. Furthermore, the large permeability of SMs allows to improve 
the critical parameters of SC wires and strips by shielding the transport current 
self-induced magnetic field as well as an externally imposed magnetic field~\cite{strip1,Campbell,wire1,wire1a,wire2,wire2a}. 
As is known from previous work, the critical current of a periodic structure of SC strips 
separated by slits is markedly enhanced compared with that of an isolated SC strip~\cite{Mawatari1}. 
Here, therefore, we study how filling the slits with SM strips controls the penetration of 
magnetic flux into such a heterostructure.

\begin{figure}[b]
%\begin{center}
\vspace{0.5cm}
\includegraphics[width=8.5cm]{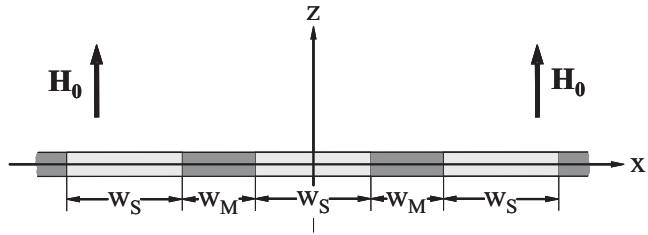}
%\end{center}
\caption{Schematic view of part of the SC/SM heterostructure composed of SC strips of width $w_S$ (light shading) 
and SM strips of width $w_M$ (dark shading). The strips, of thickness~$d$, extend infinitely in $y$-direction 
of a cartesian coordinate system $x, y, z$. The direction of the external magnetic field~${\bf H}_0$ is marked.}
\label{fig1}
\end{figure}

To this end, we consider a periodic array of infinitely extended type-II SC and SM strips of  
respective widths $w_S$ and $w_M$, {\it i.e.} period $w=w_S+w_M$, and thickness $d$, oriented parallel 
to the $x$-$y$-plane of a cartesian coordinate system $x, y, z$ and exposed to a transverse external 
magnetic field ${\bf H}_0$, as shown in Fig.~\ref{fig1}. 
The SC constituents shall be devoid of 
magnetic flux penetrated from their infinitely far ends; the SM constituents shall be fully 
described by the relative permeability $\mu$. Assuming, in addition, $d/w \ll 1$, variations 
of the current over the thickness of the SC strips may be ignored and, up to an error of order $d/w$, 
the state of these strips characterized by the sheet current $J$ alone.

The magnetic field ${\bf H}$ created by an arbitrary sheet current distribution in the heterostructure 
under consideration is conveniently decomposed according to ${\bf H}={\bf H}_S +{\bf H}_M$, where ${\bf H}_S$ 
stands for the magnetic field created by the SC strips in the absence of the SM constituents~\cite{Mawatari2} 
and  ${\bf H}_M$ denotes a magnetic correction field induced by fictitious magnetic charges distributed over 
the surfaces of the SM constituents~\cite{mcharge}.
The requirement of vanishing of the $z$-component of the total magnetic field on the surfaces $z= \pm d/2$ 
of the SC constituents, 
\begin{equation}
H_z \left(x \right) = H_0+H_{S,z} \left( x \right)+H_{M,z} \left( x \right)  
\end{equation}

\noindent with
\begin{equation}
H_{S,z} \left( x \right)= \frac{1}{2 \mu_0 w} \int_{-w_S/2}^{w_S/2} d \xi J\left(\xi\right) \cot \left[ \frac{\pi 
\left( \xi-x \right)}{w} \right]  
\end{equation}

\noindent and
\begin{eqnarray}
H_{M,z} \left( x \right) = - \frac{d}{\pi w^2} 
\frac{q}{1+q} \int_{-w_M/2}^{w_M/2} d \xi \left[ H_0+H_{S,z} \left( \xi+ \frac{w}{2} \right) \right] 
\phantom{12345678901234567890123456789}\\ 
\times  \sum_{n=0}^{\infty}  q^n \left\{ \psi' \left[ \frac{w/2+nw_M+(-1)^n \xi -x}{w} \right]
+ \psi' \left[ \frac{w/2+nw_M-(-1)^n \xi +x}{w} \right] \right\}, \nonumber
\end{eqnarray}

\noindent yields an integral equation which determines the sheet current distribution in the Meissner state.
Here, $\mu_0$ means the permeability of free space and $q=(\mu -1)/(\mu +1)$ signifies the strength of 
the image current induced by the SM constituents; $\psi$ is the digamma function.

\begin{figure}[t]
%\begin{center}
%\vspace{1.5mm}
\includegraphics[width=8cm]{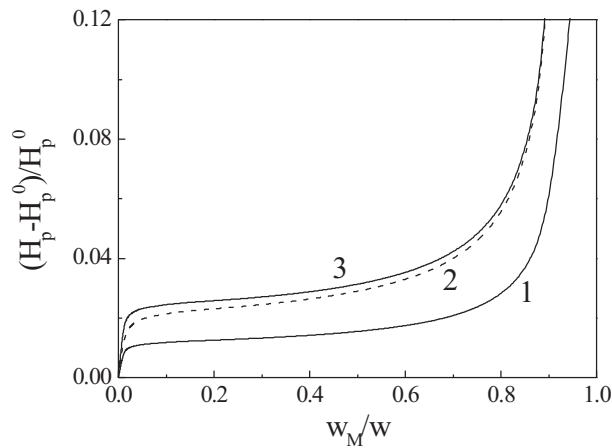}
%\end{center}
\caption{Variation of the field of complete penetration of magnetic flux into the SC strips with the width of 
the SM constituents, for $d/w=0.005$, $\mu=1000$ (curve~1), 
$d/w=0.01$, $\mu=500$ (curve~2) and $d/w=0.01$, $\mu=1000$ (curve~3).}
\label{fig2}
\end{figure}

The field of complete penetration of magnetic flux, $H_p$, corresponds to the transition of the SC 
constituents into the full critical state; it is found by setting $H_z =0$   
for $x=0$ in conjunction with the sheet current distribution $J \left( x \right)= -J_c \mathop{\rm{sgn}} x$ 
which, in turn, is governed by the critical sheet current~$J_c$. Fig.~\ref{fig2} displays the calculated 
dependence of $H_p$ on the relative width of the SM constituents using different values of $d$ and $\mu$. 
Evidently, due to the presence of even narrow SM strips, $H_p$ slightly increases compared to the 
respective field $H_p^0$ for the case of a periodic array of isolated SC strips. 
A~pronounced rise of $H_p$ occurs when the width of the SM constituents grows relative to the width of 
the SC constituents; behaviour which can be explained by the fact that, owing to their high permeability, 
the SM strips attract part of the external magnetic flux so as to decrease the effective local magnetic field 
applied to the SC constituents. This redistribution of the field also occurs in a periodic array 
of isolated SM strips, where maximum reduction of the total magnetic field halfway between the strips, 
\begin{equation}
\Delta H_z = \frac{H_0 d}{\pi w} \; \psi \left( \frac{w-w_M}{2} \right),   
\end{equation}
\noindent is attained in the limit $\mu \gg 1$.

\end{document}